# Automated generation of bacterial resource allocation models

Ana Bulović[2,$], Stephan Fischer[1,$], Marc Dinh[1], Felipe Golib[1], Wolfram Liebermeister[1,3], Christian Poirier[1], Laurent Tournier[1], Edda Klipp[2], Vincent Fromion[1,*] Anne Goelzer[1,*].

[1]INRA, UR1404, MaIAGE, Université Paris-Saclay, Jouy-en-Josas, France
[2]Theoretische Biophysik, Humboldt-Universität zu Berlin, Berlin, Germany
[3]Institut für Biochemie, Charité – Universitätsmedizin Berlin, Berlin, Germany

[$]these authors contributed equally to this work
[*]corresponding authors

# Abstract

Resource Balance Analysis (RBA) is a computational method based on resource allocation, which performs accurate quantitative predictions of whole-cell states (i.e. growth rate, metabolic fluxes, abundances of molecular machines including enzymes) across growth conditions. We present an integrated workflow of RBA together with the Python package RBApy. RBApy builds bacterial RBA models from annotated genome-scale metabolic models by adding descriptions of cellular processes relevant for growth and maintenance. The package includes functions for model simulation and calibration and for interfacing to Escher maps and Proteomaps for visualization. We demonstrate that RBApy faithfully reproduces results obtained by a hand-curated and experimentally validated RBA model for *Bacillus subtilis*. We also present a calibrated RBA model of *Escherichia coli* generated from scratch, which obtained excellent fits to measured flux values and enzyme abundances. RBApy makes whole-cell modeling accessible for a wide range of bacterial wild-type and engineered strains, as illustrated with a $CO_2$-fixing *Escherichia coli* strain.

**Availability:** RBApy is available at /https://github.com/SysBioInra/RBApy, under the licence GNU GPL version 3, and runs on Linux, Mac and Windows distributions.

# 1. Introduction

Predicting the growth behavior and internal states of cells is one of the major challenges in systems biology. Anticipating the impact of genetic or environmental perturbations would offer huge opportunities for the rational design of organisms in metabolic engineering and synthetic biology. Underpinning these goals is a more fundamental question of how cells are organized and how this organization emerged.

The concept of resource allocation between the cellular processes has recently offered a part of the answer [8,9,33,36]. Resource allocation aims to determine whether the resources available in the environment allow the cell to grow, or to survive. For instance, an enzyme needs to be sufficiently abundant to produce its metabolic flux; the metabolic network needs to provide sufficient flux of building blocks and energy to produce all enzymes, ribosomes and other cellular constituents; there need to be enough ribosomes to build all cellular proteins; and all these components need to fit into a limited cellular space. The mathematical formalization of the resource allocation problem for bacteria growing in exponential phase have led to the development of the Resource Balance Analysis (RBA) method [10,11] and its experimental validation in [12]. Other computational methods such as ME-FBA [27] and cFBA [24] that have been developed later are similar to RBA but differ in how they formalize the problem in detail. These methods predict the growth rate, the abundance of molecular machines and the associated metabolic fluxes. However, generating such models manually is an intensive and time consuming task, and so the need naturally arises for a software that would assist in their automatized generation. An ideal software should (a) help generate resource allocation models for any type of bacteria, (b) be flexible enough to describe most of the relevant cellular components and processes, (c) provide algorithms for model calibration (d) and methods for efficient simulation as well as (e) be generic enough to sustain future developments of the resource allocation framework or solvers employed.

To fulfil all of these requirements, we introduce a framework for automatically generating bacterial whole-cell resource allocation models and for their simulation in exponential growth by the RBA method [12], implemented in the open source Python package RBApy. The package requires as inputs an annotated genome-scale metabolic model (GSMM) [35]. RBApy complements the provided metabolic network with a default set of molecular machines and parameters for quick and semi-automated model generation. Using helper files, users can refine and manually curate the resulting model. Finally, we provide algorithms for model calibration when datasets (e.g. quantitative proteomics) are provided by the user. We used RBApy to generate an RBA model of *Bacillus subtilis* and validated its predictions with the hand-curated model developed and experimentally validated in [12]. We then demonstrated that a resource allocation model for the bacterium *Escherichia coli* can be generated completely from scratch, calibrated using available datasets [32,38], and that it provides good quantitative predictions even for an engineered strain [1].

# 2. Material and methods

## 2.1 RBApy package

RBApy is a Python package providing four core functionalities (Supplementary Data S1): generating RBA models from biological network and sequence data, estimating various model parameters, converting RBA models to the generic XML-rba format and converting XML-rba files into an RBA optimization problem that can be solved as described in [10,11,12]. The package runs on Linux, Mac and Windows systems and is distributed under the GNU GPL version 3 license. The complete documentation is available at https://sysbio-inra.github.io/RBApy/ (Supplementary Notes S2).

### *RBApy.prerba*: generating models from SBML and Uniprot

The primary input of RBApy is an SBML file listing the metabolites and reactions of interest, as well as linking the reactions to catalyzing enzymes, in the syntax defined by COBRA or based on the SBML fbc package for constraint-based models. RBApy automatically retrieves the latest Uniprot protein sequences for the modelled organism and generates a fully functional RBA model, which is then stored in an XML-rba files.

*RBApy.prerba* also generates a number of preconfigured helper files in TSV format, which serve as a template for the user to provide missing or additional information for a second run of *RBA.prerba*. Typical examples of missing information include enzymatic proteins that could not be retrieved in Uniprot, missing information about protein localization, or missing stoichiometrical information about the protein subunits of enzymes. The helper files provide default values or educated guesses based on Uniprot annotations that can be hand-curated to progressively enhance the model's accuracy.

*RBApy.prerba* generates a basic cell model with default molecular machines for translation and protein chaperoning. To correctly define these processes, RBApy needs the composition of ribosomes, tRNAs and chaperones. RBApy provides default FASTA files, containing information extracted for *E. coli,* that can be used to generate a first model. For more accurate models of other bacteria, the user is invited to adapt the FASTA files to the modelled organism.

On a standard laptop, the initial creation (including Uniprot querying) of the *E. coli* model took less than 30 seconds, and model updating through helper files took approximately 5 seconds (Supplementary Table S3).

This subpackage uses the libsbml, biopython and pandas libraries.

### *RBApy.xml*: maintaining models in XML format

*RBApy.prerba* is primarily designed to generate a minimal working RBA model, containing default processes such as translation and chaperoning. RBApy.xml stores these models in an XML-rba format that was designed to facilitate model extension, in particular by adding new macromolecular processes. The user may do this by changing the XML-rba files directly. Alternatively, *RBApy.xml* provides an Application Programming Interface (API) in which every XML-rba entity can be accessed through a Python class with identical name.

This subpackage uses the lxml library.

### *RBApy.core*: running simulations

*RBApy.core* imports an XML-rba model and converts it into the final LP optimization problem, specified by sparse matrices as described in [11,12]. For a given medium composition, the solver solves a series of LP feasibility problem for different growth rates, and computes in fine the maximal possible growth rate, reaction fluxes and abundances of molecular machines at maximal growth rate. The optimization problem is solved by using the CPLEX Linear Programming solver (https://www.ibm.com/analytics/cplex-optimizer). The procedure has been optimized to return results in less than one minute on a standard laptop, even for large systems such as the RBA model of *E. coli* that contains 1807 metabolites, 2583 metabolic reactions and 3906 enzyme complexes (Supplementary Table S3).

This subpackage uses the scipy and cplex libraries.

### *RBApy.estim*: estimating parameters

*RBApy.estim* offers several functions for the estimation of RBA model parameters. Based on proteomics data provided by the user (in a suitable format and with subcellular locations specified for each protein), it estimates the fraction of proteins per compartment with respect to the total protein content, either as constant numbers or as linear functions of the growth rate. The estimation procedure makes sure that these functions are always positive and sum to one. If data about protein functional assignments are available, *RBApy.estim* can provide estimates of percentage of housekeeping proteins per compartment (Supplementary Notes S4). In case the user provides proteomics and fluxomics data for the same medium, it can estimate apparent catalytic rates for a subset of enzymes.

This package uses the pandas and cvxopt libraries.

## 2.2 XML-rba format

A central question we had to tackle while developing RBApy was how to encode resource allocation models in a standard description format. In particular, we had to determine if the SBML format that is commonly used for encoding GSMM was suitable to describe all constraints and, especially, the reactions of synthesis and degradation of macromolecules that are performed by molecular machines [15]. Macromolecules can be proteins, RNAs, DNA as well as heterologous complexes composed of proteins and/or RNAs, and/or metabolites.

Let us take protein translation as an illustrative example. Translation is a cellular process that relies on ribosomes as a molecular machine and that produces proteins. During protein production, ribosomes consume and produce metabolites such as charged and uncharged-tRNAs and GTP/GDP, to cite a few. Since the amino-acid composition of proteins is known from the genome sequence, protein production could theoretically be described by metabolic-like reactions involving charged-/uncharged-tRNAs, GTP/GDP, one per protein, according to the SBML format. But specifying all these reactions explicitly for each protein species would be laborious and error-prone if we have to consider the entire cell.

Indeed, macromolecules such as proteins typically undergo several macromolecular processes (translation, chaperoning, secretion), making them effectively processed by several machines operating in series. To account for such a level of detail, one would need to define one reaction per macromolecule per process undergone, with the appropriate molecular machine such as an enzyme or a ribosome. The description of models in SBML is protein- or gene-centered, meaning that the fate of each protein has to be described explicitly by individual reactions. Therefore, if a resource allocation model contains 1000 proteins, and if all proteins require chaperones for chaperoning, then 1000 reactions for protein synthesis and 1000 reactions for chaperoning, i.e. 2000 reactions in total would have to be listed in the SBML file. The SBML file would thus grow very rapidly and would contain thousands of closely related reactions, making it extremely difficult to check the consistency of reactions or to update them upon addition of a new process. This problem would become even more severe if processes are further subdivided into smaller steps, e.g., if translation is divided into ribosome binding, chain elongation, and ribosome unbinding.

An alternative description consists in defining only one template reaction per macromolecular process, which can then produce a range of different macromolecules. This definition is process-centered instead of being protein-centered and offers a simpler and clearer view of how macromolecules are produced. This process-centered representation is generic, flexible and adaptable to a large set of bacterial cells [12,16]. Unfortunately, such template reactions cannot be encoded in the existing SBML format [41].

Consequently, for encoding resource allocation models in RBApy, we chose not to follow a protein-centered representation as in SBML, but the formal process-centered representation of the bioontology BiPON [14]. The principle behind BiPON is to break down the cell as a system into intertwined biological processes, where each biological process is described as an input/output system itself. In RBApy, a macromolecular process such as translation is described by (A) the molecular machine that catalyzes the process, (B) the list of macromolecules to be processed or produced, (C) the list of metabolites (such as GTP/GDP or (un)charged-tRNAs) consumed and/or produced by the molecular machine for functioning, and (D) the efficiency of the molecular machine in catalyzing the process. We designed an XML template that helps define the characteristics (A), (B) (C) and (D) of macromolecular processes (Supplementary Notes S2). Then, defining a new macromolecular process is straightforward and can be done by the user in a few steps (Supplementary Notes S2). The cost of all newly processed macromolecules is updated automatically. For example, for a model to cover protein secretion, translocation processes may be added. As soon as a protein is listed as an input of one of the translocation process, its overall production cost will be updated (Supplementary Figure S5). Molecular machines can be described easily by listing their individual components (proteins, RNAs, metabolites) together with their component stoichiometries.

In the XML-rba format, different types of information are stored in separate files to foster a flexible re-use of data. There are files that contain basic molecular information (metabolites, reactions, proteins, RNAs, DNA); files that describe a coupling of entities at a systemic level (enzymes and processes); and files that contain parameters (model parameters and growth medium concentrations) (Supplementary Figure S6). This separation generates flexibility because of the possible combinations (e.g. solving the same model for different catalytic rates by switching the parameter file), but also because of the formalism we use to describe cellular processes.

**Files describing biological entities (metabolites.xml, proteins.xml, rnas.xml and dna.xml).** The file metabolites.xml defines the compartments, reactions and metabolites in a simplified SBML format. The files proteins.xml, rnas.xml and dna.xml are based on a common XML format. Each file contains a list of possible macromolecule components (e.g. amino acids, vitamins and cofactors for proteins, nucleotides for RNAs) and a list of the macromolecules themselves, each being defined by its components and component stoichiometries. It is important to note that components (e.g., amino acids as building blocks of proteins) are specified independently of the metabolites (e.g., amino acids as products of the metabolic network): it is the macromolecular processes that determine how macro-components are produced from metabolites. This distinction is one of the key points that make RBApy models flexible.

**Files describing catalysis by molecular machines (enzymes.xml and processes.xml).**
The file enzymes.xml describes a coupling between reaction rates and the abundances of catalyzing enzymes or enzyme complexes and quantifies this coupling through enzymatic efficiencies. An enzyme is defined as a small molecular machine (usually a set of proteins of given stoichiometry, with or without additional cofactors) that is needed to sustain one of the metabolic fluxes. In the case of enzymes, the efficiencies correspond to the apparent catalytic rates of the enzyme. For an irreversible enzyme, we use only one value. For reversible enzymes, we use two values for the apparent catalytic rates for the forward and reverse directions. Mathematically, for a reversible enzyme with concentration E, catalyzing a reaction with apparent catalytic rates $k_{app,\leftarrow}$ and $k_{app,\rightarrow}$, the reaction flux $v$ is constrained by the inequalities

$$-k_{app,\leftarrow}E \leq v \leq k_{app,\rightarrow}E$$

This constraint defines a minimal concentration $E$ necessary to sustain a given flux $v$ in forward or backward direction and, as a consequence, minimal production fluxes for the macromolecules composing the enzyme, to balance their dilution by cell growth (see [11-12] for more information).

In the metabolic network, a reaction can be catalyzed by several distinct isoenzymes, and a single enzyme can catalyze several reactions. Since each isoenzyme has its own enzyme kinetics, isoenzymes have different apparent catalytic rates. In the model, isoenzymes are treated as separate enzymes, catalyzing separate (yet identical) reactions, in order to ensure a one-to-one mapping between a reaction and an enzyme with a unique apparent catalytic rate. In the same way, a multifunctional enzyme has necessarily different affinities with respect to the different substrates, and so different apparent catalytic rates. Multifunctional enzymes are thus duplicated by introducing different apparent catalytic rates, one (or two if the reaction is reversible) per reaction. The total amount of a multifunctional enzyme is obtained by summing up the individual enzyme amount per reaction.

The file processes.xml describes a coupling between macromolecule components and metabolites by macromolecular processes. Each process is performed by a molecular machine with an efficiency, defined similarly as an enzyme efficiency, and is specified by (i) a map between component and metabolites called ProcessingMap (Supplementary Notes S2), and (ii) a list of input macromolecules. The production reaction of each macromolecule is obtained by combining the ProcessingMaps of all processes in which it appears as an input (Supplementary Figure S5).

**Cell description files (density.xml and targets.xml).** The file density.xml contains bounds on the maximal macromolecule density in each compartment, expressed in millimoles of amino acid residues per gram of cell dry weight [mmol.AA/gCDW]. Every macromolecule contributes to this density with a numerical weight defined as the sum of its components' weights expressed in mmol.AA/gCDW; the solver ensures that the total macromolecule concentration must not exceed the bounds defined in density.xml.

The file targets.xml defines fluxes and concentrations of compounds that must be maintained for the cell to be functional, such as key metabolite renewal, maintenance ATP production or the abundances of housekeeping proteins. The numerical values that fluxes and concentrations of compounds must match are specified in the file parameters.xml (see below). The module *RBApy.core* will automatically generate an equality constraint to assign the fluxes and concentrations of compounds to their specified values. Three types of targets can be considered: concentrations (keeping a metabolite or macromolecule at a given concentration), component fluxes (producing or degrading a metabolite or macromolecule at a given rate), and target reaction fluxes (forcing a reaction to occur at a given rate).

**Parameters (parameters.xml).** The file parameters.xml contains all numerical values occurring in the model (except for reaction and component stoichiometries). These are: total amino acid concentrations, fractions of protein per compartment, percentages of non-enzymatic protein per compartment, target fluxes and concentrations for metabolites and macromolecules, efficiencies of all molecular machines. Each parameter is defined as a constant value or as a function of growth rate or external metabolite concentrations. Supported function types are: linear, inverse, exponential and Michaelis-Menten. Typically, enzyme efficiencies vary linearly with growth rates, and transporters have efficiencies that depend on the concentration of metabolites transported. Alternatively, a parameter can be defined as a combination of the basic functions. The dependency of parameters with respect to variables such as growth rate can thus be defined easily. For instance, maximal macromolecule densities can be defined either as constant or growth rate dependent.

## 2.3 Parameter estimation from existing datasets

**Global parameters.** RBA models contain different types of parameters. A set of parameters that globally restrain the cell state are the cytosolic density and the growth-rate dependent total concentration of proteins, both expressed in millimoles of amino-acid residues per gram of cell dry weight [mmol.AA/gCDW]. To estimate these two parameters, we consolidated data available from three sources [3,23,25], choosing the most comprehensive dataset, which proved to be the one from [3] (Supplementary Notes S7). We estimated the maximal cytosolic density for a growth rate of ~1 [1/h], for which all three datasets provide information (albeit of a different kinds) and for which two data sources match almost perfectly, thereby obtaining the value of 4.89 mmol.AA/gCDW (see Supplementary Notes S7 for detailed procedures). Additionally, we used proteomics data measured for different growth rates [32] to estimate the percentage of proteins allocated to individual compartments (extracellular, periplasm, cytoplasm, inner plasma membrane and outer plasma membrane). The estimated total protein concentration and the percentage of proteins allocated to individual compartments were used to compute the maximal protein density per compartment.

**Enzyme efficiencies.** We calculated estimates of the individual apparent catalytic rates of enzymes for the growth on glucose minimal medium at 37°C, for which fluxomics and proteomics data of comparable growth rates (< 5% difference) and experimental setups were available [32,38]. We chose not to use the measurements done on galactose because the difference in growth rates between fluxomics and proteomics was quite big (~30%), which might imply a significant difference in cellular configurations. To obtain complete and consistent flux distributions for the estimation, we ran an FBA simulation [40] of the original metabolic model, with flux constraints representing the experimentally measured fluxes. Apparent catalytic rates were computed as $k_{app_i} = v_i/E_i$, when $v_i$ and $E_i$ do not vanish, and where $v_i$ is taken from the resulting flux distribution and $E_i$ is the enzyme concentration computed from proteomics data.

**Other molecular machine efficiencies.** An RBA model requires parameters defining the efficiencies of molecular machines of the macromolecular processes. Using proteomics data from 12 different media [32], we estimated the molecular machine efficiencies for chaperoning and secretion as linear functions of the growth rate as in [12] (Supplementary Notes S7). For each of the twelve different media (and thus twelve different growth rates) we computed the efficiency of a molecular machine as the ratio between the production flow of molecules by the process at measured growth rate and the measured abundance of the molecular machine. In the estimation procedure, the measured abundance of the molecular machine (e.g. chaperones and the secretion apparatus) is computed from proteomics data for each medium. The production flow of molecules by the process (e.g. the flow of folded and secreted proteins) at measured growth rate is estimated through the constraint C$_2$. For chaperoning, the flow of folded proteins is computed as $\beta\mu^{mes}P_{tot}(\mu^{mes})$ where β is the fraction of total proteins that are folded by a chaperone and is equal to 10% [5], and the total concentration of proteins $P_{tot}$ is the one estimated previously. For secretion, we assumed that proteins that are located either in the inner membrane ($P_i$), periplasm ($P_p$), outer membrane ($P_o$) or in the medium ($P_e$) need to be translocated by the secretion apparatus. The flow of secreted proteins is thus computed as $\mu^{mes}\left(P_i(\mu^{mes}) + P_p(\mu^{mes}) + P_o(\mu^{mes}) + P_e(\mu^{mes})\right)$. The efficiencies of the molecular machines are then fitted by a linear function of the growth rate with the coefficient of determination for chaperoning of $R^2 = 0.97$ and for secretion of $R^2 = 0.98$.

# 3. Results

## 3.1 A pipeline to generate, refine, calibrate and simulate RBA models

In Figure 1, we summarize the cell description underlying the RBA method and the mathematical relationships defining the interactions and allocation of resources between the cellular processes [10,11,12]. All these relationships take the form of linear growth-rate dependent equalities and inequalities (Figure 1): for cells growing in exponential phase at a rate $\mu$, (I) the metabolic network has to produce all metabolic precursors necessary for biomass production (equalities $C_1$ in green); (II) the capacity of all molecular machines must be sufficient to ensure their function, i.e. to catalyze chemical conversions at a sufficient rate (inequalities $C_2$ in blue for the enzymes and transporters, in yellow for the molecular machines of macromolecular processes); (III) the intracellular density of compartments and the occupancy of membranes are limited (inequalities $C_3$ in orange); (IV) mass conservation is satisfied for all molecule types (equalities $C_1$ in green). Taken together, the equalities and inequalities define, at a given rate $\mu$, a feasibility linear programming (LP) problem that can be solved efficiently [26]. Parsimonious resource allocation between cellular processes is modelled mathematically by maximizing the cell growth, and computed by solving a series of such LP feasibility problems for different growth rate values [10-12]. For a given medium, solving an RBA optimization problem predicts the maximal possible growth rate, the corresponding reaction fluxes and the abundances of molecular machines. Consequently, generating an RBA model requires information for formalizing constraints $C_1$, $C_2$ and $C_3$ (Figure 1), and in particular: (i) the localization and the composition of the molecular machines, (ii) the molecules that are consumed and released by the molecular machines for functioning; (iii) the efficiencies of molecular machines, i.e. the rates of the process per amount of the catalyzing molecular machine; (iv) other parameters such as the maximal density of each compartment. The number of parameters per molecular machine and per compartment, and the sources of information needed to identify those parameters, are given in Figure 2. We introduced one (resp. two) efficiency per irreversible (resp. reversible) enzyme and transporter, one efficiency per molecular machine performing a macromolecular process, and two parameters (the maximal density and the fraction of house-keeping proteins) per compartment as described in [12]. The final number of parameters in the RBA model depends on the type of function (e.g. constant, linear, Michaelis-Menten) and of the variables (e.g. growth-rate, extracellular concentrations of nutrients) used to model the molecular machine efficiencies, on the maximal density, and on the fraction of housekeeping proteins per compartment (Figure 2). The choice of function and of the variables used is up to the modeller, and may depend on the biological phenomenon to be modelled and on the quantity of datasets available for model calibration. For instance, in [12] we used four datasets, each of them composed of fluxomics and quantitative proteomics, that were acquired in four growth conditions. The four growth conditions were chosen so as to obtain a wide range of growth rates to estimate linear functions of enzyme efficiencies. Estimating the maximal density and the fraction of housekeeping proteins per compartment requires quantitative proteomics data as well as information about protein localisation and about the belonging of each protein to the set of housekeeping proteins (Figure 2). Details on the procedure of parameter estimation from datasets are given below.

### 3.1.1 Pipeline description

We designed RBApy, a free and open-source Python package, to assist RBA model generation and simulations (Supplementary Data S1, Supplementary Notes S2). RBApy is composed of three main modules (Figure 3): ***preRBA*** for generating of RBA models represented in an XML format, ***core*** for simulation and visualization of the results, and ***estim*** for parameter estimation.

As inputs, the user needs to provide a GSMM in Systems Biology Markup Language (SBML) format [15], as well as the NCBI taxonomy identifier of the bacterium being modelled. The GSMM must contain gene annotations for each enzyme-catalyzed reaction. The first pipeline module, *RBApy.preRBA,* automatically downloads the latest Uniprot protein annotations of the modelled organism [37]. Then it (1) extracts the metabolic network and the enzymatic proteins associated with each annotated metabolic reaction from the GSMM, and retrieves (2) the sequence, localization, subunit stoichiometry and cofactors of enzymatic proteins from their Uniprot annotations. The lack of biological databases readily offering information of macromolecular processes such as translation or chaperoning (Figure 1) prevents the complete automatized generation of RBA models. The same holds true for certain model parameters, such as the maximal density of the cytosol. To complete the missing information, we offer a default set of parameters and macromolecular processes (translation and chaperoning) of *E. coli.* From all the collected information, *RBApy.preRBA* generates the resource allocation model encoded in a specific XML format, called XML-rba. The XML-rba format is based on SBML and was conceived to provide an efficient description of cellular processes. The model obtained in this first run can already be simulated. Users can then refine the default macromolecular processes (1) by providing correct sequence information for molecular machines in FASTA format and (2) by manually editing the precise chemical formulae of reactions associated to macromolecular processes in the XML-rba files. Alongside the model in XML format, *RBApy.preRBA* generates several helper files which contain information that remained ambiguous (such as cofactor IDs that couldn't be automatically linked to model metabolites) and propose default values. The user can choose either to use default values, either to update them manually. The RBA model can be iteratively refined by editing the helper files, by specifying new macromolecular processes, or by directly editing the XML-rba files, and then re-running *RBApy.preRBA* to generate the final model.

If the user provides data sets containing quantitative proteomics and fluxomics data in comparable growth conditions, the module *RBApy.estim* can be used to estimate the main types of model parameters [12]: maximal macromolecular compound densities within cellular compartments, abundance of housekeeping proteins for each compartment, and the efficiencies of molecular machines involved in metabolic and macromolecular processes (Figure 1).

The *RBApy.core* module starts by converting the XML-rba files into an RBA convex optimization problem where growth rate is maximized by running a series of LP feasibility problems [12]. Results can be exported in various formats suitable for visualizing the fluxome and proteome. If the reactions in the original SBML file carry BIGG identifiers [18], flux distributions can be visualized via Escher maps [17], a web-based tool for pathway visualization. Individual protein abundances and the cost of cellular processes can be visualized as Proteomaps [21] or pie charts using user-defined or COG functional annotations [34].

### 3.1.2 Validation of the implementation of RBA constraints within RBApy

We illustrate the progressive improvements of models, using helper files, in Supplementary Data S8. We used RBApy to generate an RBA model of *B. subtilis* semi-automatically in three stages, and compared each time the predicted growth rate and flux distribution to the values from an RBA hand-curated and experimentally validated model [12]. During the first stage of model generation, RBApy could not automatically match some of metabolite identifiers between SBML and RBApy (Supplementary Figure S9A), resulting in an overestimated growth rate value of 2.36 [1/h] in glucose minimal medium. The manual matching of identifiers improved prediction accuracy (Supplementary Figure S9B). After model calibration, the final model shows an excellent quantitative agreement with the original hand-curated model (growth rate = 0.64 [1/h], $R^2 \geq 0.99$, Supplementary Figure S9C). The remaining discrepancies between the fluxes in different model versions are due to, for example, a different usage of enzymatic cofactors (imported from Uniprot during the first run). The complete list of differences is given in Supplementary Data S8.

## 3.2 A resource allocation model for an *Escherichia coli* wild-type strain

RBApy can also be used to model bacteria with several compartments and to estimate their parameters. As an illustrative example, we have selected the Gram-negative model bacterium *E. coli*, which has two compartments: the cytoplasm and the periplasm (Supplementary Data S10). As a metabolic network reconstruction, we chose the iJO1366 model [29], which was the most up-to-date genome-scale metabolic reconstruction at the time of model generation. To supplement the metabolic reconstruction, we retrieved protein sequence information from the Uniprot database [37] (done automatically through RBApy), as well as the DNA, rRNA, tRNA and mRNA sequences from the NCBI database [7] (done manually).

### 3.2.1 Parameter estimation from existing datasets

We estimated the model parameters from three sources [3,23,25] and datasets [32,38] as described in Material and methods (see Supplementary Notes S7 for further details). Among the model parameters, the efficiency of enzymes, also called "apparent catalytic constant" ($k_{app}$) describes the reaction rate per amount of the catalyzing enzyme and can be directly compared to values reported in the literature. A default value of 12.5 [1/s] is first set for all enzymes. This value was chosen since it maximized the coefficient of determination ($R^2$) of a linear fit between predicted and measured growth rates on 12 different media [32] (Supplementary Figure S11), and is remarkably close to the median of reported maximal turnover rates of enzymes, 10 [1/s] [2,32]. We additionally estimated the individual apparent catalytic rates of 417 enzymes for growth on glucose minimal medium, for which fluxomics and proteomics data with comparable growth rates (< 5% difference) and experimental setups were available [32-38]. We compared the statistical distribution of the computed apparent catalytic rates to the distribution of maximal turnover rates of *E. coli* enzymes reported in the BRENDA database [30]. As shown in Figure 5A, the distributions span roughly the same range of values. A perfect match is not expected, considering the many possible sources for discrepancies. First, apparent catalytic rates are expected to be lower than the maximal turnover rates. Moreover, most of the values reported in BRENDA stem from *in-vitro* measurement, in many differing measurement conditions, which may differ strongly from *in vivo* conditions [12]. We also compared our predictions to those of [4] (Supplementary Figure S12) and obtained a good general match across 184 enzymes.

### 3.2.2 Validation of the predictions of the RBA model

We validated the predictions for two model versions. First, we used only the default apparent catalytic rate of 12.5 [1/s] for all enzymes. Indeed, this case is representative of the use of RBA models when no datasets are available for model calibration. We ran simulations of the *E. coli* RBA model for 12 different media for which measured growth rates and proteomics were available for exponential growth in batch culture from [32]. Except for the specified carbon sources and the added amino acids, the minimal medium assumed in the model was the one specified for the iJO1366 metabolic reconstruction [29]. As can be seen in Figure 4, the predicted growth rates show a good match to the measured ones ($R^2 = 0.54$), even if the default apparent catalytic rate was used.

Because flux and proteomics data were available for growth on glucose, allowing us to estimate the apparent catalytic rate of 417 enzymes, we took a deeper look at model behaviour with this carbon source. We ran a simulation of the *E. coli* RBA model using the estimated $k_{app}$ values on a glucose minimal medium (see Supplementary Figure S13 and S14 for the visualization of results on Escher Maps [17] on Proteomaps [21]). We then compared the predicted enzyme amounts to the enzyme abundance estimates from proteomics data, and we performed this comparison for both model versions, before and after the enzyme-specific parameter estimation. In the predictions from the first model, the one with default catalytic rate of 12.5 [1/s], the predicted protein levels do not correlate well with measurements, with the membrane and cofactor biosynthesis enzymes being far off from the measured values (Figure 5C). Using the $k_{app}$ estimation values greatly improved the predictions, with the correlation for 233 cytosolic enzymes being $R^2 = 0.65$ in log space (Figure 5D). In an RBA model, the growth medium is defined by listing the concentrations of extracellular metabolites. The transporter efficiencies are computed from these concentrations by assuming a Michaelis-Menten transporter kinetics. The precise estimation of parameters of this Michaelis-Menten kinetics requires additional datasets describing the kinetics of consumption of the extracellular metabolites. Whenever such data were not available, we took as default values the ones estimated for *B. subtilis* transporters. Their abundances and corresponding fluxes are predicted from simulations. In the case of a glucose minimal medium with $k_{app}$ estimates, the glucose uptake and acetate excretion fluxes were predicted almost exactly, with the overall flux distribution matching the measured one with $R^2 = 0.89$ (Figure 5B).

### 3.2.3 Prediction of complex cell behaviour

In RBA, each metabolic flux is supported by an enzyme which, in turn, must be produced in its exact macromolecular composition, including cofactors, from precursors produced in metabolism. In spite of this circular dependence between metabolism and enzymes, RBA can predict complex behaviour, such as the choice between isoenzymes. To investigate this, we focused on the usage of different dehydrogenases in the *E. coli* respiratory chain depending on the presence of a certain cofactor in the medium. *E. coli* is equipped with many dehydrogenases that can be part of the respiratory chain. One of them, the pyrroloquinoline quinone-dependent D-glucose dehydrogenase [39], is relatively cheap in terms of protein cost compared to other dehydrogenases, but it requires a cofactor called pyrroloquinoline quinone (PQQ) for which no *de novo* biosynthetic pathway exists in *E. coli*. The model predicts that in glucose minimal medium, where this cofactor is absent, *E. coli* will use the NADH dehydrogenase, while upon addition of the PQQ in the medium, it will use the D-glucose dehydrogenase instead, in agreement with experiments [39], and grow faster. The predicted flux distributions (Supplementary Figure S13, S15) indicate that this change occurs together with other metabolic rearrangements, such as activation of the pentose-phosphate pathway and another substitution, that of transhydrogenase. This example illustrates the accuracy of RBA predictions with respect to the exact composition of the medium and the predictive capacity provided by the concept of parsimonious resource allocation between cellular processes.

## 3.3 A resource allocation model for engineered *Escherichia coli* cells

To show the general applicability and flexibility of the RBA modelling pipeline, we adapted our wild-type *E. coli* model, based on the metabolic network reconstruction *iJO1366*, to an engineered *E. coli* strain capable of $CO_2$ fixation, developed by [1]. To adjust our model, we introduced four novel reactions: two catalyzed by the type II Rubisco enzyme (from *Rhodospirillum rubrum ATCC 11170*), one by a phosphoribulokinase (from *Synechococcus elongatus PCC 7942*) and one by a carbonic anhydrase (*from Rhodospirillum rubrum*), each with their proper amino acid and ion composition. At the same time, we removed the two reactions of the glyoxylate shunt (MALS and ICL), two of glycolysis (PFK and PGM) and one PPP reaction (G6PDH2r) to model the gene deletions in the carbon-fixing strain designed in [1]. We additionally removed one other reaction of the glyoxylate metabolism (GLYCK), which produces 3-phospho-D-glycerate, but is only induced on glycolate as the carbon source [28].

The final engineered strain grows slowly at a rate of 0.12 [1/h] [1], far below the growth rate of 0.65 [1/h] in glucose minimal medium. The metabolic state of the engineered strain is, thus, expected to differ quite strongly from the one of the wild-type strain growing in glucose. The apparent catalytic rates of the engineered strain should thus be quite different from the ones that were estimated in glucose minimal medium. Therefore, we preferred to use again a default value of 12.5 [1/s] for apparent catalytic rates of all enzymes except for those of carbon fixation. We modelled Rubisco activity using either a Michaelis-Menten function of the $CO_2$ concentration as $\frac{k_{max}CO_2}{K_M+CO_2}$, or as competitive inhibition by oxygen as $\frac{k_{max}[CO_2]}{K_M(1+[O_2]/K_I)+[CO_2]}$ assuming that $CO_2$ is dissolved in water under 0.1 atm and $O_2$ is present and normal atmospheric conditions and room temperature (Table 1). The parameter values used were taken from Brenda database for *Rhodospirillum rubrum* or as median, minimal and maximal values as reported in [6]. The efficiency of the *prk* enzyme was set to 12.5 [1/s], the default $k_{app}$ value for *E. coli*. For the carbonic anhydrase efficiency, we used the median of the values reported in Brenda for that enzyme, which is ~10000 [1/s]. The updated model is provided in Supplementary Data S16.

The new model, incorporating all the specified knock-outs and recombinantly expressed proteins, yields a growth rate in the range of 0.16 to 0.18 [1/h], which is close to the experimental one of 0.12 [1/h] [1], with Rubisco taking up from ~1 to ~18% of cytosolic proteins (Table 1), depending on the model type and on the kinetic parameter values chosen to describe Rubisco activity. The changes in resource allocation between different cellular functions can be seen on Supplementary Figure S17. Upon removal of $CO_2$ from the medium, the simulated cell is no longer viable, confirming that $CO_2$ is an indispensable carbon source for this engineered strain. RBApy offers unique ways of investigating the CO2-fixing *E. coli* model, especially through the possibility to screen parameters easily and to visualize the impacts of parameter changes on predictions on Escher maps and Proteomaps. Refined with dedicated datasets, the current CO2-fixing *E. coli* model may constitute a first stage of a whole-cell model-assisted strain design for synthetic biology.

# 4. Discussion

RBApy relies on the systemic description of the cell [14,16,19,20], centered around the notion of cellular processes, an efficient way to describe the interconversion of any kinds of compounds in the cell. We chose a flexible way to define cellular processes by listing (1) the composition of the molecular machine that catalyzes the process, (2) a list of macromolecules to be processed or produced by the molecular machine, (3) a list of metabolites consumed and/or produced by the molecular machine for functioning, and (4) the efficiency of the molecular machine in catalyzing the process. We show that the main cellular processes (i.e. translation, chaperoning and secretion of proteins) of two bacterial species can be properly handled using this description. If transporter, enzyme, or machine efficiencies vary depending on growth rate or extracellular nutrient concentrations, this can be described using constant, linear, exponential or Michaelis-Menten functions. For instance, the activity of RubisCO of the $CO_2$-fixing *E. coli* strain was assumed to depend on the extracellular $CO_2$, and the relationship was modelled by a Michaelis-Menten function with or without a competitive inhibition of the extracellular $O_2$. Other cellular processes with other types of efficiency parameters or other dependencies, for example on temperature, can be easily integrated. RBApy was designed to be as flexible as possible. We made sure that new solvers and new algorithms for parameter calibration or result visualization can be easily integrated.

Generating a new RBA model necessitates only little information: a genome-scale metabolic model in SBML format [15], the composition of the molecular machines, and the NCBI taxon ID to download protein information from Uniprot [37]. While RBApy collects most information automatically, manual curation (through helper files) can be necessary to remove ambiguities in annotations or to extend the model by adding other macromolecular processes. To limit the need for manual curation, SBML files should be more systematically annotated using the protocol of [35]. Cross-references to other databases, such as the ChEBI IDs [13] for metabolites or Uniprot IDs for proteins, should be present systematically. Furthermore, existing standard formats of Systems Biology are currently too limited to encode whole-cell models and especially to support DNA, RNA, and protein sequence-based reaction patterns [41]. During RBApy development, we faced the same difficulties, which led us to develop the XML-rba format. We showed, through the models RBA-Bsub and RBA-Ecoli, that the XML-rba format is generic enough to describe (i) the major bacterial macromolecular processes in exponential growth easily, and (ii) resource allocation models. Future versions of the format will be updated in agreement with the progresses of the systems biology community at refining formats to account for whole-cell descriptions.

RBApy complements ongoing initiatives for the generation of resource allocation models [22,31] by demonstrating, for the first time, that a calibrated resource allocation model for a new bacterium can be generated completely from scratch. Calibrating the models for new organisms requires proteomics data, but with the increasing availability of quantitative proteomics, fluxomics, and exchange fluxes data, missing model parameters will cease to be a major limitation in a near future. RBApy thus makes whole-cell modelling and simulation accessible for a large diversity of prokaryotes, and future releases of RBApy will cover also eukaryotic cells, cell communities and simulations in dynamical conditions.

# Acknowledgments

This work has received funding from the French Lidex-IMSV of Université Paris-Saclay and from the People Programme (Marie Skłodowska-Curie Actions) of the European Union's Horizon 2020 Programme under REA grant agreement no. 642836. This material reflects only the authors' views and the Union is not liable for any use that may be made of the information contained therein. We also thank O. Inizan for testing the pipeline.

# Author contributions

AG and VF conceived and supervised the study. SF, AB, AG and FG conceived and implemented RBApy. SF conceived the XML-rba format and generated the RBA-Bsub model, with the help of MD, LT and AG. AB generated the two RBA-Ecoli models with the help of EK, WL and AG. AB, CP and AG developed the website and the virtual machine. AB, SF, WL, AG and VF wrote the paper with inputs from all authors.

# Competing interest

The authors declare no competing interest.

# Figure Legends

**Figure 1.** Resource Balance Analysis models. A growing cell is described by a network model comprising metabolic and macromolecular processes. Model variables (compound concentrations and reaction fluxes) are linked by physical and biochemical constraints, describing mass balances, the relation between reaction fluxes and catalyst concentration, and upper limits on total molecule concentrations. These constraints give rise to a tractable, growth-rate dependent linear feasibility problem. For a given growth medium, the maximal possible growth rate and associated model variables are computed by solving a series of feasibility problems for different growth rate values.

**Figure 2.** Number of parameters per molecular process and per compartment in a RBA model, and the type of data necessary for parameter estimation.

**Figure 3.** Pipeline for generating an RBA model from an existing genome-scale metabolic model and from the NCBI Taxon ID of the modelled organism.

**Figure 4.** Predictions of *E. coli* growth rates (in the order of increasing growth rate) for galactose, acetate, pyruvate, fumarate, succinate, glucosamine, glycerol, mannose, xylose, glucose, fructose minimal media and for glycerol with 20 amino acids. Dots show results based on a default apparent catalytic rate of 12.5 [1/s] for all enzymes. The 'x' marker denotes the prediction for growth on glucose using the estimated apparent catalytic rates. The coefficient of determination between estimated and measured growth rates is $R^2 = 0.58$.

**Figure 5.** Parameter estimation and model predictions on glucose minimal medium. **(A)** Cumulative histogram of estimated apparent catalytic rates of 452 enzymes in glucose minimal medium and of catalytic rates of *E. coli* reported in Brenda database. **(B)** Comparison of measured and predicted flux values for CCM fluxes, as well as glucose uptake and acetate excretion flux, from a model using estimated $k_{app}$ values. **(C)** Comparison of measured abundances of cytosolic enzymes to predictions from a model with default enzyme $k_{app}$ values and **(D)** a model using estimated $k_{app}$ values.

# Table legends

**Table 1**: Predictions of growth rate and percentage of Rubisco in cytosol for *E. coli* $CO_2$ fixing strain for two different enzyme kinetics, Michaelis-Menten with $CO_2$ as substrate and competitive inhibition by oxygen. All values are computed for $CO_2$ at 0.1 atm (experimental condition in [1]) and $O_2$ at standard atmospheric conditions.

[a]Growth rate expressed as log(2)/$T_d$, with $T_d$ being the doubling time.

[b]Percentage of Rubisco computed in terms of amino acids assigned to Rubisco compared to total amino acid content in the cytoplasm.

[c]Michaelis-Menten enzyme kinetics $\frac{k_{max}[CO_2]}{K_M+[CO_2]}$.

[d]Competitive inhibition by $O_2$ kinetics: $\frac{k_{max}[CO_2]}{K_M(1+[O_2]/K_I)+[CO_2]})$

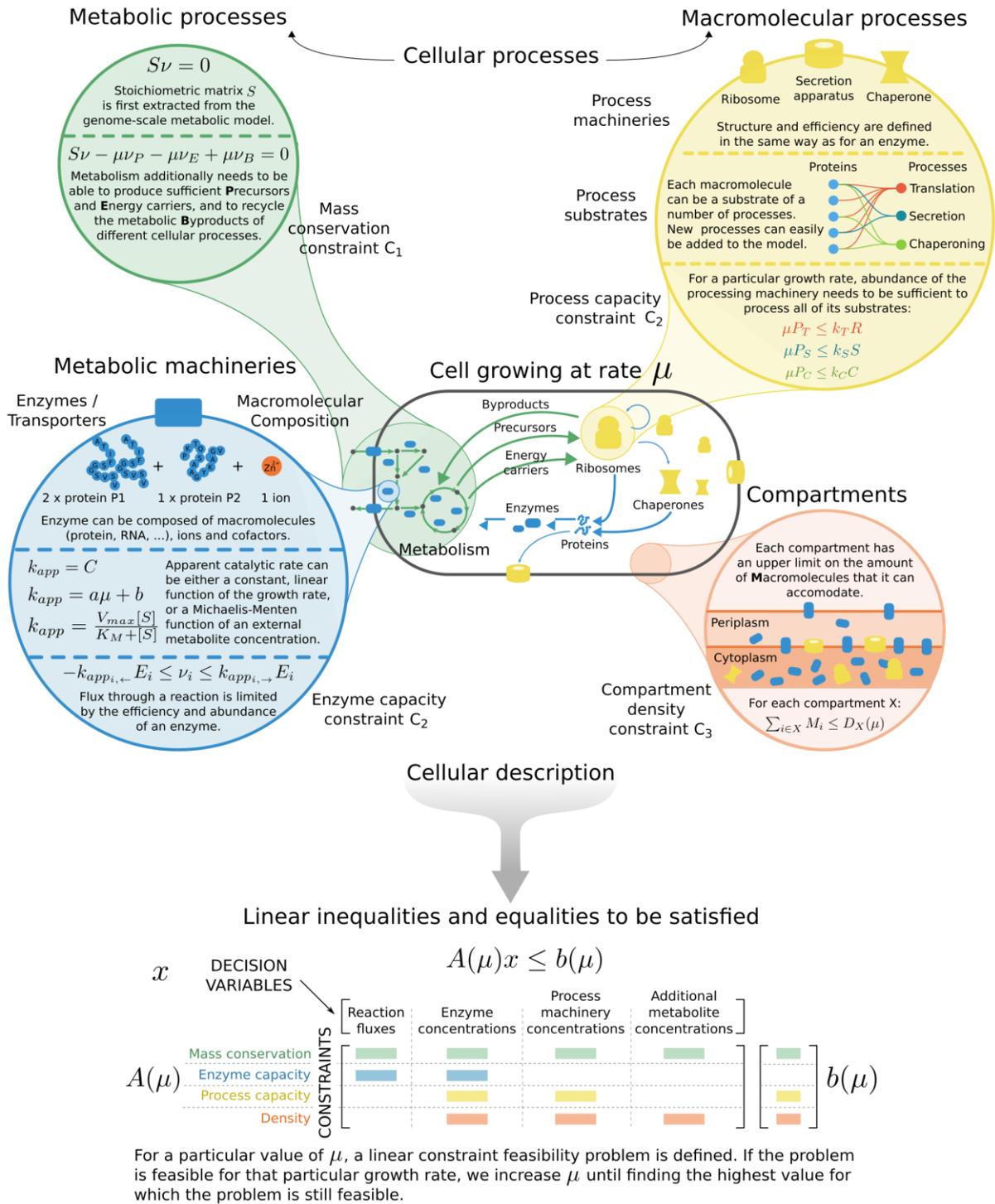

**Figure 1**

|  | | Parameter description | Parameteric function | Functions per unit | Count in *E. coli* model | Experimental data |||
|---|---|---|---|---|---|---|---|---|
|  | | | | | | P | F | W* |
| MOLECULAR MACHINE EFFICIENCY | METABOLISM | Irreversible enzyme | constant | 1 | 1891 | X | X | |
| | | Reversible enzyme | | 2 | 450 | | | |
| | | Irreversible transporter | MM | 1 | 138 | X | X | |
| | | Reversible transporter | MM, constant | 2 | 1095 | | | |
| | CELLULAR PROCESS | Macromolecular process machine efficiency | linear (μ) | 1 | 3 | X | | |
| COMPARTMENT | | Compartment density | linear (μ) | 1 | 5 | X | | X |
| | | % of housekeeping protein | linear (μ) | 1 | 5 | | | |

*P - proteomics, F - fluxomics, W - cell weight and macromolecular composition measurements

| Function type | Parameter count | Parameter type |
|---|---|---|
| constant | 1 | constant |
| linear | 2 | slope, intercept |
| Michaelis Menten | 2 | $k_{max}, K_M$ |

Function variables:
 - Growth rate ($\mu$)
 - Concentration of an external metabolite

| | |
|---|---|
| Original GSMM reaction number | 2583 |
| RBA reaction number (isoreactions included) | 3574 |
| Total parameter number | 6378 |
| Total parameter number (using default enzyme and transporter efficiencies) | 31 |

**Figure 2**

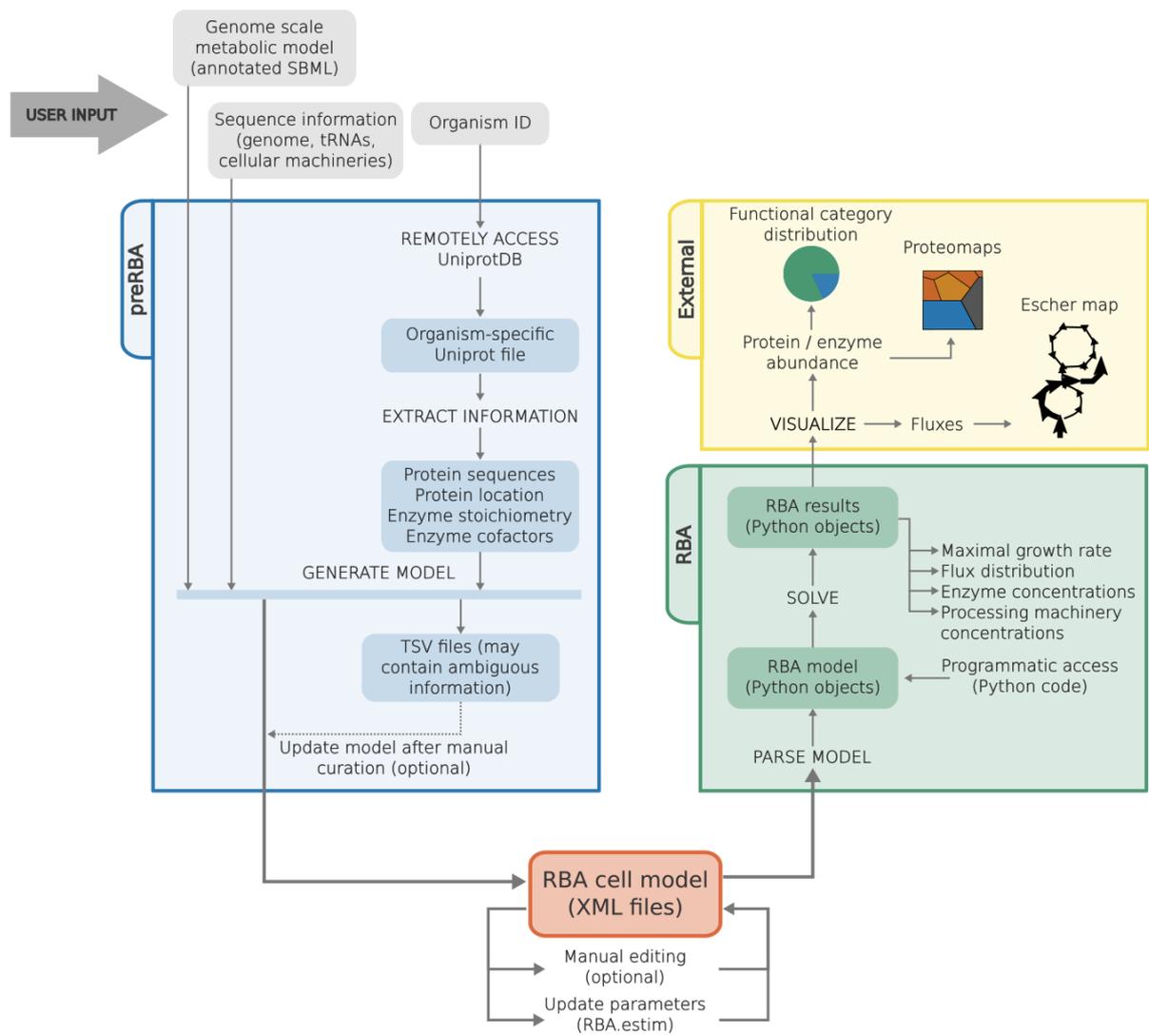

**Figure 3**

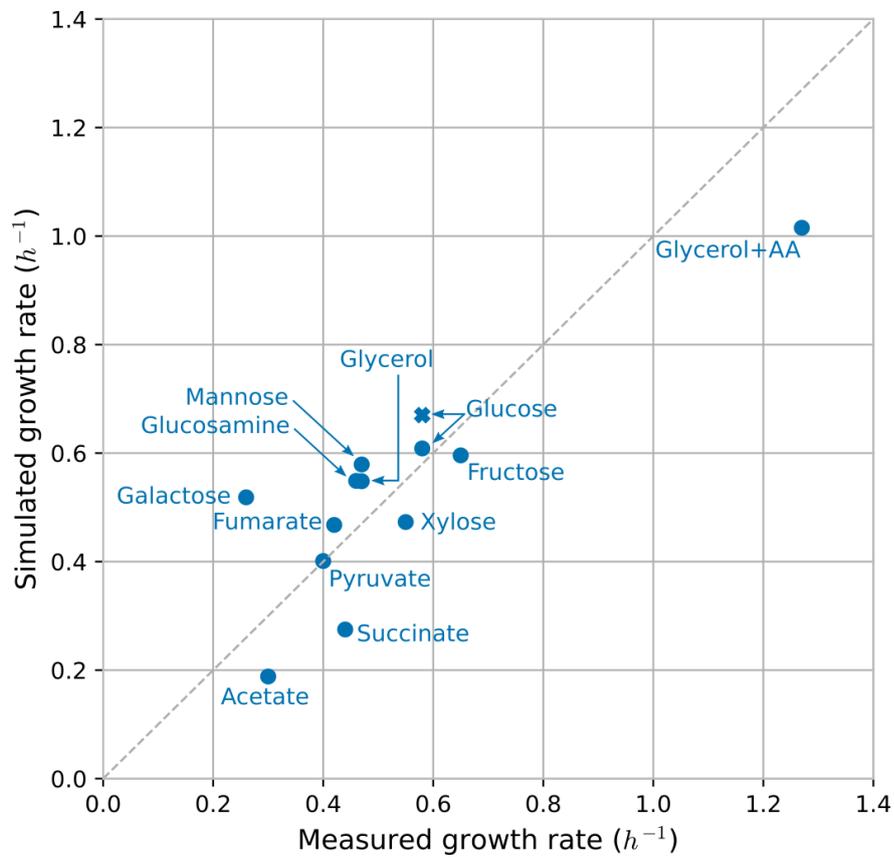

**Figure 4**

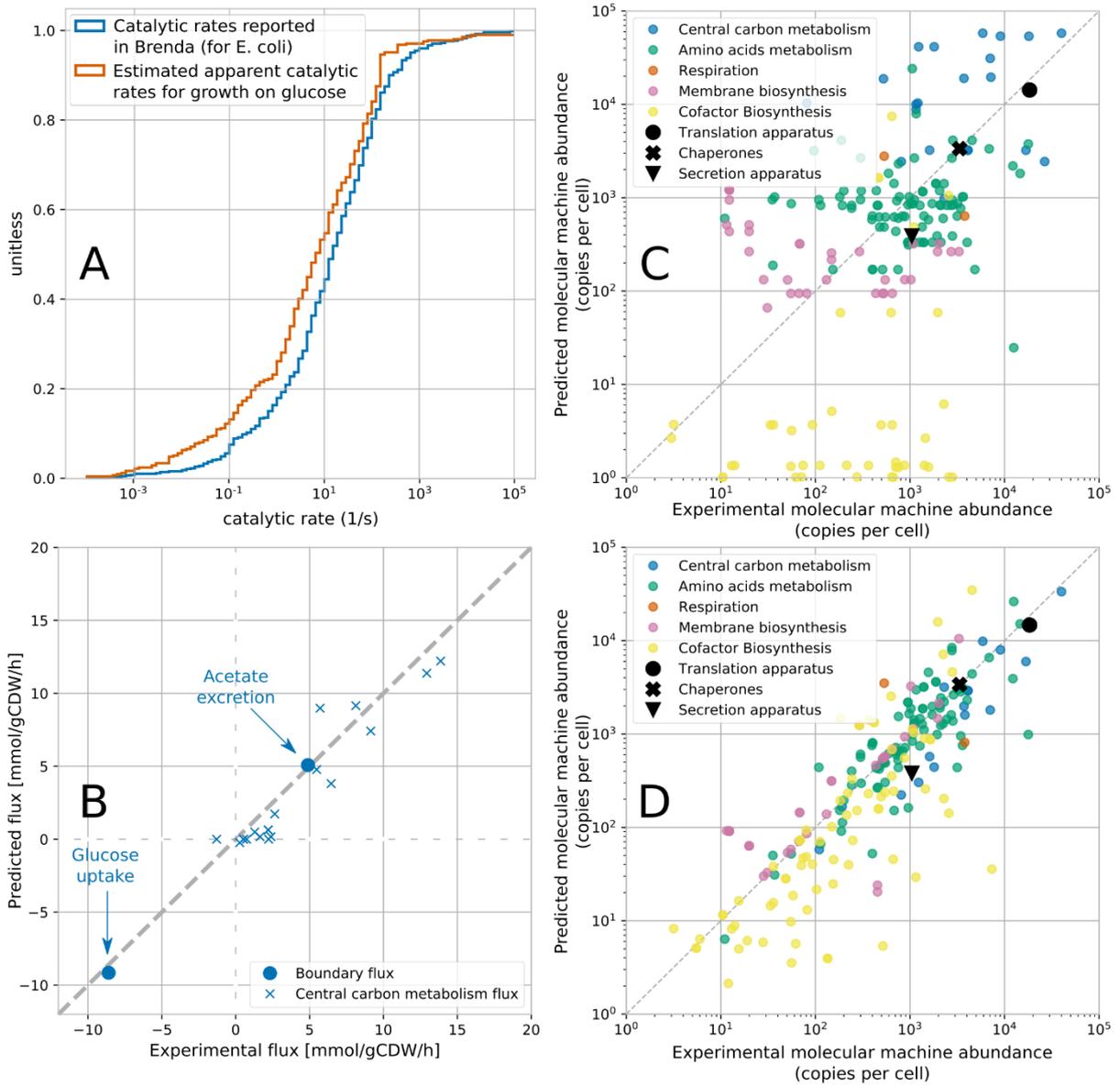

**Figure 5**

| Enzyme kinetics of Rubisco | | | Growth rate [1/h][a] | % of Rubisco in cytosol[b] | Comment |
|---|---|---|---|---|---|
| Michaelis-Menten with $CO_2$ as substrate[c] | | | | | |
| $k_{max}$ [1/s] | $K_M$ [µM] | | | | |
| 1.31 | 446 | | 0.1827 | 4.98 | Median values taken from Brenda for *R. rubrum* |
| 1.31 | 14 | | 0.183 | 4.44 | $K_M$ set to median $K_C$ value of [6] |
| 0.32 | 14 | | 0.1685 | 17.7 | $k_{max}$ set to lowest $k_{cat,C}$ value of [6] |
| 12.6 | 14 | | 0.1866 | 0.54 | $k_{max}$ set to highest $k_{cat,C}$ value of [6] |
| Competitive inhibition by $O_2$[d] | | | | | |
| $k_{max}$ [1/s] | $K_M$ [µM] | $K_I$ [µM] | | | |
| 3.16 | 14 | 446 | 0.1854 | 1.88 | Median of values of [6] for $k_{cat,C}$, $K_C$ and $K_O$ |
| 0.32 | 14 | 446 | 0.17 | 16.11 | $k_{max}$ set to lowest $k_{cat,C}$ value of [6] |
| 12.6 | 14 | 446 | 0.1867 | 0.48 | $k_{max}$ set to highest $k_{cat,C}$ value of [6] |

**Table 1**